\documentclass[twocolumn,showpacs, amsmath, amssymb, floatfix,
                nofootinbib,wrapfloat byrevtex]{revtex4}
 \usepackage{pstricks,pst-plot,pst-coil,pstricks-add}
 \usepackage{graphicx,epstopdf}
 \usepackage[bookmarks=true,%
  bookmarksopen=true,%
  pdfborder={0 0 0},%
  pdfhighlight={/N},%
  linkbordercolor={.5 .5 .5}]{hyperref}

 \newcommand{\beq}[1]{\begin{eqnarray}\label{#1}}
 \newcommand{\eeq}{\end{eqnarray}}
 \makeatletter
 \def\dPlotEbar{\pst@object{dPlotEbar}}
 \def\dPlotEbar@i#1{\begin@SpecialObj\expandafter\dPlotEbar@GetCoordinate#1}
 \def\dPlotEbar@GetCoordinate#1{\pst@cntc=1\dPlotEbar@DoCoordinate#1}
 \def\dPlotEbar@DoCoordinate(#1,#2,#3,#4){
 \ifnum#1=2
 \psline[linecolor=black](!#2 #3 #4 0.5 mul add)(!#2 #3 #4 0.5 mul sub)
 \psdot[linecolor=black,dotstyle=triangle](#2,#3)
 \else
 \psline[linecolor=cyan](!#2 #3 #4 0.5 mul add)(!#2 #3 #4 0.5 mul sub)
 \psdot[linecolor=cyan,dotscale=0.4](#2,#3)
 \fi
 \advance\pst@cntc by 1 %
 \@ifnextchar({\dPlotEbar@DoCoordinate}{\end@SpecialObj}
 }
 \makeatother

 \def\dataLatticeGG{(1,0.347534,-1.163340,0.000729)(1,0.491483,-0.874449,0.000729)(1,0.601961,-0.743813,0.000729)(1,0.695066,-0.714751,0.000729)(1,0.777119,-0.631593,0.000729)(1,0.851282,-0.576925,0.000729)(1,0.982962,-0.508730,0.001458)(1,1.042600,-0.489450,0.001458)(1,1.042600,-0.477075,0.001458)(1,1.203890,-0.406580,0.002188)(1,1.390130,-0.340975,0.002188)(1,1.474450,-0.306733,0.002188)(1,1.554200,-0.280549,0.002917)(1,1.702560,-0.231633,0.003646)(1,1.737670,-0.224151,0.002917)(1,1.805850,-0.197967,0.005104)(1,1.965960,-0.152791,0.004375)(1,2.085200,-0.121427,0.004375)(1,2.085200,-0.118550,0.005834)(1,2.331320,-0.053808,0.005834)(1,2.407780,-0.028774,0.008750)(1,2.432730,-0.027048,0.005834)(1,2.457440,-0.020430,0.005834)(1,2.553850,0.006618,0.007292)(1,2.780260,0.063591,0.008021)(1,2.948940,0.105026,0.009480)(1,3.009730,0.125168,0.013855)(1,3.108440,0.140706,0.009480)(1,3.127800,0.149626,0.009480)(1,3.127800,0.150201,0.013126)(1,3.405120,0.211203,0.014584)(1,3.440400,0.219548,0.012396)(1,3.475340,0.233647,0.011667)(1,3.611660,0.263572,0.018959)(1,3.822860,0.313351,0.014584)(1,3.885530,0.318531,0.014584)(1,3.931880,0.330616,0.016772)(1,4.170400,0.390179,0.016772)(1,4.170400,0.395070,0.021147)(1,4.213630,0.401975,0.026980)(1,4.256390,0.411761,0.022605)(1,4.423370,0.442260,0.021876)(1,4.517940,0.467581,0.020418)(1,4.662650,0.488586,0.023334)(1,4.815560,0.555056,0.040106)(1,4.865460,0.542105,0.026980)(1,4.914850,0.560521,0.029897)(1,5.107650,0.614041,0.032814)(1,5.213000,0.617494,0.033543)(1,5.213000,0.641091,0.035731)(1,5.406340,0.678207,0.038648)(1,5.417510,0.693458,0.061253)(1,5.439760,0.662383,0.036460)(1,5.560540,0.691158,0.043752)(1,5.897850,0.782369,0.048127)(1,5.958950,0.794744,0.053232)(1,6.019440,0.850852,0.091879)(1,6.216850,0.816613,0.055419)(1,6.255600,0.897466,0.061982)(1,6.389330,0.908401,0.062711)(1,6.621410,1.000480,0.129798)(1,6.810210,0.993575,0.087504)(1,6.880820,1.026090,0.081670)(1,7.223340,1.129960,0.173550)(1,7.298200,1.090260,0.115214)(1,7.372300,1.100610,0.112297)(1,7.825290,1.146360,0.248657)(1,7.863760,1.244200,0.158966)(1,8.340800,1.217720,0.196155)(1,8.427220,1.244200,0.329598)(1,9.029190,1.286780,0.487835)(1,9.631120,1.334550,0.689823)(1,0.475641,-0.941468,0.001066)(1,0.672651,-0.694642,0.000533)(1,0.823857,-0.578590,0.001066)(1,0.951280,-0.549784,0.001066)(1,1.063580,-0.473046,0.001066)(1,1.165080,-0.420277,0.002131)(1,1.345300,-0.352577,0.002131)(1,1.426920,-0.332814,0.002131)(1,1.426920,-0.319569,0.003197)(1,1.647670,-0.245774,0.004795)(1,1.902560,-0.175973,0.003730)(1,2.017950,-0.138130,0.004795)(1,2.127110,-0.107854,0.004795)(1,2.330160,-0.051720,0.007992)(1,2.378200,-0.043941,0.006394)(1,2.471520,-0.013456,0.011722)(1,2.690650,0.039946,0.009590)(1,2.853850,0.078211,0.009590)(1,2.853850,0.082836,0.014386)(1,3.190700,0.163359,0.012787)(1,3.295340,0.188798,0.023976)(1,3.329470,0.193634,0.013320)(1,3.363300,0.202043,0.015984)(1,3.495240,0.234421,0.020246)(1,3.805120,0.309898,0.018648)(1,4.035960,0.353629,0.025042)(1,4.119180,0.368766,0.041558)(1,4.254260,0.405558,0.028238)(1,4.280770,0.411023,0.036763)(1,4.660310,0.502061,0.044755)(1,4.708590,0.497646,0.042091)(1,4.943010,0.554831,0.073526)(1,5.381250,0.653016,0.068198)(1,5.707680,0.702001,0.107093)(1,5.766850,0.719660,0.130003)(1,6.590670,0.856949,0.230702)(1,0.256189,-1.438810,0.001768)(1,0.362304,-1.085390,0.003674)(1,0.443730,-0.934326,0.007249)(1,0.512377,-0.902245,0.003758)(1,0.572854,-0.807876,0.004051)(1,0.627529,-0.743900,0.005232)(1,0.724610,-0.664353,0.007627)(1,0.768563,-0.649283,0.005277)(1,0.768563,-0.638104,0.007088)(1,0.887461,-0.563824,0.010713)(1,1.024750,-0.500590,0.007342)(1,1.086910,-0.462953,0.010355)(1,1.145710,-0.437406,0.008090)(1,1.255060,-0.389913,0.011618)(1,1.280940,-0.378757,0.008039)(1,1.331190,-0.365271,0.012972)(1,1.449220,-0.305635,0.011241)(1,1.537130,-0.286571,0.010410)(1,1.537130,-0.275821,0.010938)(1,1.718560,-0.224156,0.009832)(1,1.774920,-0.216985,0.018587)(1,1.793320,-0.204912,0.010114)(1,1.811520,-0.211200,0.014447)(1,1.882590,-0.174255,0.012694)(1,2.049500,-0.128777,0.012640)(1,2.173830,-0.094107,0.016903)(1,2.218650,-0.066900,0.021229)(1,2.291410,-0.051696,0.017519)(1,2.305690,-0.053597,0.013989)(1,2.305690,-0.042945,0.017029)(1,2.510120,-0.004910,0.017922)(1,2.536140,0.020547,0.020451)(1,2.561890,0.010496,0.015162)(1,2.662390,0.044557,0.023983)(1,2.818070,0.074554,0.017107)(1,2.864270,0.089563,0.016782)(1,2.898430,0.105415,0.025191)(1,3.074250,0.137887,0.018497)(1,3.074250,0.138102,0.022364)(1,3.106110,0.159036,0.034303)(1,3.137660,0.135877,0.022691)(1,3.260750,0.193987,0.030272)(1,3.330440,0.183362,0.020950)(1,3.437130,0.226115,0.023254)(1,3.549850,0.291879,0.035799)(1,3.586640,0.260075,0.022535)(1,3.623040,0.268978,0.034737)(1,3.765180,0.297309,0.030348)(1,3.842830,0.314944,0.024255)(1,3.842830,0.327548,0.029038)(1,3.985350,0.364478,0.036847)(1,3.993590,0.399596,0.045554)(1,4.009980,0.346726,0.027783)(1,4.099010,0.390693,0.027338)(1,4.347640,0.476203,0.040516)(1,4.392700,0.444316,0.035889)(1,4.437300,0.479956,0.058956)(1,4.582820,0.490792,0.031004)(1,4.611380,0.534695,0.038361)(1,4.709960,0.501462,0.043880)(1,4.881040,0.573713,0.064948)(1,5.020250,0.605249,0.047108)(1,5.072280,0.584829,0.048282)(1,5.324760,0.650486,0.076873)(1,5.379950,0.663709,0.047686)(1,5.434570,0.689690,0.060137)(1,5.768500,0.809368,0.083061)(1,5.796880,0.778775,0.066235)(1,6.148500,0.873617,0.064934)(1,6.212240,0.822039,0.094757)(1,6.655960,0.890356,0.112580)(1,7.099700,1.087550,0.123936)(2,1.267110,-0.383459,0.006100)(2,1.520530,-0.291715,0.007600)(2,1.773950,-0.210348,0.008900)(2,2.027370,-0.136203,0.009900)(2,2.280790,-0.066753,0.011000)(2,2.787630,0.064254,0.011000)(2,3.041050,0.126206,0.012000)(2,3.294480,0.187369,0.013000)(2,3.547900,0.248138,0.014000)(2,3.801320,0.309301,0.024000)}
 \savedata{\rhoEqqbAnaAprxMone}[{{0.05,-2.88716},{0.065,-2.41964},{0.1,-1.83118},{0.15,-1.42698},{0.2,-1.20356},{0.25,-1.05753},{0.3,-0.952196},{0.35,-0.871056},{0.4,-0.805502},{0.45,-0.750565},{0.5,-0.703143},{0.55,-0.661175},{0.6,-0.623205},{0.65,-0.58815},{0.7,-0.55514},{0.75,-0.523427},{0.8,-0.492293},{0.85,-0.451558},{0.9,-0.414086},{0.95,-0.379362},{1.,-0.346974},{1.05,-0.316587},{1.1,-0.28793},{1.15,-0.260776},{1.2,-0.234938},{1.25,-0.210257},{1.3,-0.1866},{1.35,-0.163854},{1.4,-0.14192},{1.45,-0.120715},{1.5,-0.100166},{1.55,-0.0802094},{1.6,-0.0607894},{1.65,-0.0418575},{1.7,-0.0233705},{1.75,-0.00529029},{1.8,0.012417},{1.85,0.0297816},{1.9,0.0468305},{1.95,0.0635881},{2.,0.0800762},{2.,0.0800762},{2.5,0.233692},{3.,0.373996},{3.5,0.506691},{4.,0.634633},{4.5,0.759404},{5.,0.881956},{5.5,1.0029},{6.,1.12262},{6.5,1.24142},{7.,1.35949},{7.5,1.47697},{8.,1.59397}}]

 \savedata{\chispdofMone}[{{5,0.134576},{4.5,0.150296},{4,0.150166},{3.5,0.165051},{3,0.149043},{2.5,0.147075},{2,0.154300},{1.5,0.262064},{1,1.679330},{0.92,5.415},{0.5,29.847548},{0,45.264567}}]
 \savedata{\chispdofMtwo}[{{5,0.136408},{4.5,0.147692},{4,0.151695},{3.5,0.165300},{3,0.150302},{2.5,0.144278},{2,0.146986},{1.5,0.213268},{1,1.342383},{0.915,5.43683},{0.5,29.740786},{0,42.137851}}]
 \savedata{\chispdofMthree}[{{5.0,0.137084},{4.5,0.154001},{4.0,0.150427},{3.5,0.168122},{3.0,0.149979},{2.5,0.145249},{2.0,0.151685},{1.5,0.256044},{1.0,1.256620},{0.853,5.43},{0.5,28.652265}}]

 \savedata{\chispdofMone}[{{5,0.134576},{4.5,0.150296},{4,0.150166},{3.5,0.165051},{3,0.149043},{2.5,0.147075},{2,0.154300},{1.5,0.262064},{1,1.679330},{0.92,5.415},{0.5,29.847548},{0,45.264567}}]
 \savedata{\chispdofMtwo}[{{5,0.136408},{4.5,0.147692},{4,0.151695},{3.5,0.165300},{3,0.150302},{2.5,0.144278},{2,0.146986},{1.5,0.213268},{1,1.342383},{0.915,5.43683},{0.5,29.740786},{0,42.137851}}]
 \savedata{\chispdofMthree}[{{5.0,0.137084},{4.5,0.154001},{4.0,0.150427},{3.5,0.168122},{3.0,0.149979},{2.5,0.145249},{2.0,0.151685},{1.5,0.256044},{1.0,1.256620},{0.853,5.43},{0.5,28.652265}}]

 \savedata{\rhoEqqbListaMone}[{{0.005465,-6.272562},{0.010175,-4.163911},{0.016654,-3.038731},{0.024872,-2.363258},{0.034673,-1.918955},{0.045909,-1.607484},{0.058368,-1.376893},{0.071860,-1.198940},{0.086236,-1.057303},{0.101297,-0.941423},{0.116933,-0.844190},{0.133035,-0.761164},{0.149455,-0.689148},{0.166195,-0.625631},{0.183139,-0.568939},{0.200229,-0.517863},{0.217409,-0.471256},{0.234691,-0.428216},{0.252085,-0.388439},{0.269566,-0.351197},{0.287007,-0.316256},{0.304492,-0.283115},{0.322065,-0.251547},{0.339757,-0.221565},{0.357442,-0.192675},{0.375342,-0.164541},{0.393274,-0.137121},{0.411430,-0.110247},{0.429693,-0.084443},{0.448262,-0.058477},{0.466971,-0.033041},{0.486097,-0.007642},{0.505573,0.017812},{0.525371,0.043103},{0.545671,0.068871},{0.566651,0.094783},{0.588170,0.121192},{0.610614,0.148544},{0.633872,0.176269},{0.658307,0.205162},{0.683854,0.235095},{0.710777,0.266574},{0.739554,0.299981},{0.770288,0.335314},{0.803669,0.373865},{0.840699,0.415711},{0.881678,0.462332},{0.927802,0.514918},{0.981834,0.575793},{1.046734,0.648689},{1.128331,0.740899},{1.241882,0.868127},{1.429076,1.078848},{1.973,1.6885},{2.224695,1.963761}}]
 \savedata{\rhoEqqbListaMtwo}[{{0.005401,-6.910736},{0.010117,-4.565931},{0.016637,-3.321848},{0.024937,-2.578363},{0.034870,-2.091121},{0.046284,-1.750575},{0.058964,-1.499095},{0.072721,-1.305441},{0.087393,-1.151620},{0.102789,-1.025980},{0.118787,-0.920701},{0.135270,-0.830945},{0.152092,-0.753198},{0.169250,-0.684659},{0.186624,-0.623558},{0.204154,-0.568568},{0.221781,-0.518424},{0.239517,-0.472170},{0.257370,-0.429439},{0.275314,-0.389463},{0.293218,-0.351984},{0.311169,-0.316461},{0.329210,-0.282633},{0.347372,-0.250524},{0.365527,-0.219601},{0.383901,-0.189515},{0.402307,-0.160196},{0.420938,-0.131477},{0.439682,-0.103913},{0.458734,-0.076187},{0.477929,-0.049036},{0.497549,-0.021928},{0.517526,0.005218},{0.537828,0.032178},{0.558645,0.059647},{0.580155,0.087258},{0.602215,0.115394},{0.625221,0.144521},{0.649058,0.174042},{0.674096,0.204789},{0.700273,0.236656},{0.727855,0.270151},{0.757334,0.305682},{0.788815,0.343278},{0.823002,0.384275},{0.860926,0.428769},{0.902890,0.478345},{0.950118,0.534233},{1.005441,0.598957},{1.071888,0.676437},{1.155426,0.774457},{1.271675,0.909678},{1.463311,1.133612},{1.973,1.73226},{2.277800,2.089699}}]
 \savedata{\rhoEqqbListaMthree}[{{0.003207,-9.708400},{0.006938,-5.738472},{0.012562,-3.895090},{0.020125,-2.887360},{0.029540,-2.268312},{0.040621,-1.853850},{0.053180,-1.558900},{0.066972,-1.338757},{0.081825,-1.167465},{0.097516,-1.029695},{0.113901,-0.915517},{0.130868,-0.819766},{0.148169,-0.737187},{0.165852,-0.664904},{0.183844,-0.600988},{0.201926,-0.543791},{0.220226,-0.491801},{0.238566,-0.444407},{0.257080,-0.400664},{0.275571,-0.359637},{0.294073,-0.321465},{0.312782,-0.285398},{0.331392,-0.251224},{0.350157,-0.218569},{0.368901,-0.187011},{0.387784,-0.156801},{0.406944,-0.127579},{0.426004,-0.098860},{0.445313,-0.070883},{0.464837,-0.043240},{0.484694,-0.015957},{0.504787,0.010883},{0.525184,0.037784},{0.546022,0.064537},{0.567327,0.091642},{0.589148,0.118912},{0.611487,0.146587},{0.634746,0.174978},{0.658629,0.203807},{0.683779,0.233503},{0.709799,0.264077},{0.737182,0.296561},{0.766442,0.330230},{0.797215,0.365888},{0.830194,0.404385},{0.866535,0.445299},{0.906226,0.490428},{0.950080,0.540727},{1.000087,0.596982},{1.058396,0.662374},{1.128481,0.741209},{1.216790,0.840643},{1.339064,0.977502},{1.541479,1.204191},{1.973,1.67571},{2.399701,2.103140}}]

 \savedata{\FtvCurModela}[{{0.890616,7.98787},{0.845599,5.25892},{0.783347,3.52954},{0.695751,2.41188},{0.566484,1.67616},{0.343778,1.18343}}]
 \savedata{\FtvCurModelb}[{{0.889552,6.97599},{0.863774,5.46775},{0.833272,4.34232},{0.797239,3.49085},{0.754621,2.83833},{0.703969,2.33223},{0.643136,1.93531},{0.568615,1.62073},{0.473729,1.36897},{0.341617,1.16564}}]
 \savedata{\FtvCurModelc}[{{0.894165,6.69035},{0.873402,5.46375},{0.849767,4.51675},{0.82296,3.77604},{0.79262,3.18968},{0.758297,2.72033},{0.719414,2.34073},{0.675188,2.03077},{0.624509,1.7754},{0.565687,1.56326},{0.495895,1.38564},{0.409657,1.23585},{0.292941,1.10867}}]
 \savedata{\FtvSimplea}[{{0.862865,6.76758},{0.83286,5.27017},{0.798006,4.16493},{0.757483,3.33483},{0.710173,2.7016},{0.654456,2.21179},{0.587795,1.82813},{0.505715,1.52416},{0.398616,1.28082},{0.234098,1.08417}}]
 \savedata{\FtvSimpleb}[{{0.862865,6.76758},{0.83286,5.27017},{0.798006,4.16493},{0.757483,3.33483},{0.710173,2.7016},{0.654456,2.21179},{0.587795,1.82813},{0.505715,1.52416},{0.398616,1.28082},{0.234098,1.08417}}]
 \savedata{\FtvSimplec}[{{0.862865,6.76758},{0.83286,5.27017},{0.798006,4.16493},{0.757483,3.33483},{0.710173,2.7016},{0.654456,2.21179},{0.587795,1.82813},{0.505715,1.52416},{0.398616,1.28082},{0.234098,1.08417}}]

\begin{document}

\title{Heavy Quark Potentials in Some Renormalization Group Revised AdS/QCD Models}

\author{Ding-fang Zeng}
\email{dfzeng@bjut.edu.cn}

\affiliation{Institute of Theoretical physics,
Beijing University of Technology}

\begin{abstract}
We construct some AdS/QCD models by the systematic procedure of GKN.
These models reflect three rather different asymptotics
the gauge theory beta functions approach at the infrared region,
$\beta\propto-\lambda^2, -\lambda^3$ and $\beta\propto-\lambda$, where $\lambda$ is the 't Hooft
coupling constant. We then calculate the heavy quark potentials in these models by
holographic methods and find that they can more consistently fit
the lattice data relative to the usual models
which do not include the renormalization group improving effects.
But only use the lattice QCD heavy quark potentials as constrains, we cannot
distinguish which kind of infrared asymptotics is the better one.
\end{abstract}

\date{\today}
\maketitle


 \section{Introduction}
The Duality between string theories in
Anti-deSitter space and conformal field theories on
its boundary, i.e., AdS/CFT,
are powerful tools for understanding the strong coupling
gauge theory phenomena. In its most well known formulations,
the duality is between the 4-dimensional $\mathcal{N}=4$ super
Yang-Mills field theory and type IIB super string theory
in the $AdS_5\times S_5$ background \cite{Maldacena:1997re},
\cite{GKPWitten:1998}. In practice, our greatest
interest is not $\mathcal{N}=4$ SYM, which is conformal
and non-confining, instead it is the
$N_c=3$, $N_f=4$ or $N_f=6$ Quantum Chromodynamics, i.e. QCD, which has
running coupling and confining properties.

To obtain predictions about the practical QCD theories through
AdS/CFT, or more generally, the holographic principle, people
develop two ways, (i) the top-down method and (ii) the
bottom-up approach. In the first method, people either construct
dual string/gravity descriptions for gauge theories with
running couplings, e.g. cascading gauge theories\cite{CascadingGaugeTheory},
or build models for gauge theories with
fewer super-symmetries relative to the $\mathcal{N}=4$
SYM\cite{nonCon-AdSCFT-model}\cite{SakaiSugimoto}. For examples,
some models implement fundamental flavors by adding probe branes
in various exact or asymptotically $AdS_5\times X_5$ background
\cite{FlavoredAdSCFT}\cite{backreacted-flavor}, see also \cite{gaugeGravityMesonReview0711}
for reviews. The top-down
method preserves fundamental structures of string theory and its dual
gravity description is 10 dimensional.

In the second way, now known as AdS/QCD, people start from some
5 dimensional effective gravity theory and calculate the
relevant 4D gauge theory quantities through holographic method. By
requiring the results coincide with those from QCD phenomenologies,
people obtain the general properties of the dual gravity background,
which can be refined for further phenomenological goals.
In the earliest implementation of this ideal by Polchinsky and Strassler
\cite{AdSQCD1}, an infra-red cut off on the
dual $AdS_5$ background is introduced to implement confinement,
hence the name hard wall model. Since the space-time of gravity
descriptions ends below the cut off, appropriate boundary conditions
must be imposed on this point by hand, see e.g., reference \cite{BrodskyTeramond}
and \cite{RoldPmarol0501}. In reference
\cite{EKSS}, it is proposed that, by adding quark
masses/condensates, this model can become a remarkably good
model of chiral symmetry broken dynamics. Reference
\cite{dilatonDeform-adsQCD} considered a natural extension of this model
by replacing the pure AdS background with a dilaton-deformed one and get very
similar results. To improve the Regge behavior of highly excited rho mesons and higher spin mesons in
the hard wall model, reference \cite{KKSS} proposed the soft wall model, where
the infra-red hard wall is replaced by a dilaton
quadratically depending on the holographic radial
coordinate. While reference \cite{AndreevZakharov}
and \cite{Shuryak0605} proposed to include the same effects by
a simple factor $e^{cz^2}$ in front of the dual geometry's metric. Different from
this guess and trial approach, in references \cite{BCCKP0505},
\cite{GKN0707-1324} and \cite{GKN0707-1349} GKN(G\"ursoy, Kiritsis, and Nitti)
propose a systematic procedure for dual geometry's
construction and provide several models as illustrations.

We will introduce
GKN's procedure and construct our own models in the next section.
To make our dual geometries as simple as possible, so
that more QCD quantities can be calculated conveniently and analytically,
we do not require our models' $\beta$ function to
exactly coincide with the perturbative QCD theory to higher loops, like
those provided in GKN's original work. The next next section
discusses the calculation of heavy
quark potentials in the resulting models. Numerical fittings
as well as discussions about the quantum corrections are also
provided in that section. The last section
contains our main conclusions and prospects for future studies.

\section{The Models}\label{sectionModelBuilding}

By GKN's arguments, the non-critical string background dual to
QCD like gauge theories can be described by the following
Einstein frame action
 \beq{}
 S=\frac{1}{16\pi G_5}\int
 d^5x\sqrt{-G}\Big[R-\frac{4}{3}(\partial\Phi)^2+V(\Phi)\Big]
 \label{actionGravityDilaton}
 \eeq
where we have neglected the effects of an axion field, which is
of $\mathcal{O}(N_c^{-2})$ relative to the terms
written out explicitly; the potential $V(\Phi)$ encodes various
contributions such as the higher $\alpha'$ corrections and the
integration of RR 4-form field whose flux seeds the $D3$ branes and
$U(N_c)$ gauge group. Its functional form cannot be derived out explicitly
from the first principle, but can be determined through inputs
of QCD phenomenologies, e.g., the $\beta$ function describing
the running of coupling constant.

Setting the dual gravity geometry as
 \begin{subequations}
 \beq{}
 \Phi&=&\phi(u)\\
 ds^2&=&e^{2A}(-dt^2+d\vec{x}\cdot d\vec{x})+du^2
 \label{domainWall-bgrnd},
 \eeq
 \end{subequations}
and introducing phase space variable $X=\frac{\Phi'}{3A'}$, GKN
find that the dilaton field's equation of motion and Einstein equations
following from the action \eqref{actionGravityDilaton} can be
written as first order forms,
 \begin{subequations}
 \beq{}
 \Phi'&=&\frac{\sqrt{3V_0}}{2}X
 e^{-\frac{4}{3}\int_{-\infty}^\Phi Xd\Phi}
 \\
 A'&=&\frac{\sqrt{3V_0}}{6}
 e^{-\frac{4}{3}\int_{-\infty}^\Phi Xd\Phi}
 \\
 V(\Phi)&=&V_0(1-X^2)e^{-\frac{8}{3}\int_{-\infty}^\Phi Xd\Phi}
 \label{VofPhiGeneral}
 \eeq
 \label{eom-using-Xvariable}
 \end{subequations}
where $\Phi'=\frac{d\Phi}{du}$ and $A'=\frac{dA}{du}$.
After identifying the exponentiated dilaton field with
the 't Hooft coupling constant $\lambda\equiv N_cg_{YM}^2$
and the metric function $e^{A(u)}$ with the
QCD energy,
 \beq{}
 e^\Phi=\lambda\;,\;e^{A(u)}\propto E
 \eeq
GKN set up an explicit relation between the gauge theory
$\beta$ function and the gravity theory quantities $\Phi$ and $A$,
 \beq{}
 \beta(\lambda)\equiv\frac{d\lambda}{d\ln E}
 =\lambda\frac{d\Phi}{dA}=3\lambda X
 \label{GKN-X-beta-relation}
 \eeq
To implement asymptotical freedom in the dual gauge
theory, it is required that $A(u)\rightarrow u$, $\Phi\rightarrow-\infty$ as
$u\rightarrow\infty$, i.e., the dual space-time is asymptotically
anti-deSitter.
Obviously, in this construction of AdS/QCD models,
the non-perturbative $\beta$ function determines the whole structure of
the dual gravity theories. In turn, as long as we know the latter,
we can use it to calculate more QCD phenomenological quantities
through the holographic method.

In this paper, we will consider 3 models, each with
exact (non-perturbative effects included) $\beta$ function
 \beq{}
 \mathrm{Model\;1\;:\;}
 &&\beta(\lambda)=-b\lambda^2
 \label{betaFunction1}
 \\
 \mathrm{Model\;2\;:\;}
 &&\beta(\lambda)=-b_0\lambda^2-b_1\lambda^3
 \;,\;b_0\;,\;b_1>0
 \label{betaFunction2}
 \\
 \mathrm{Model\;3\;:\;}
 &&\beta(\lambda)=-\frac{b_0\lambda^2}{1+b_1\lambda}
 \;,\;b_0\;,\;b_1>0
 \label{betaFunction3}
 \eeq
respectively.
At the perturbative limit, model 1 reproduces QCD's beta
function to 1 loop order; model 2 reproduces the desired
$\beta$ function to 2 loop order; model 3 only
reproduces the desired(perturbative) $\beta$ funtion to 1 loop order.
Nevertheless, there are evidences that non-perturbative effects indeed
produce $\beta$ functions of the model 3 type, see e.g. \cite{InfraredRunningCoupling1}
,\cite{InfraredRunningCoupling2} and \cite{InfraredRunningCoupling3}. The
author of reference \cite{InfraredRunningCoupling3} tells us that,
this kind of beta function reproduces exactly all of the known results and
also the two loops. At the strong coupling
limit, model 1's $\beta$ function approaches infinite as
$-\lambda^2$, model 2 as $-\lambda^3$, while model 3, $-\lambda$.
By GKN's general confinement criteria, model 3 is critical confining,
while model 1 and model 2 are super-confining.

For model 1, according to GKN's procedure,
 \beq{}
 \beta=-b\lambda^2=\frac{d\lambda}{dA}
 \;\Rightarrow\; A=\frac{1}{b\lambda}
 \label{AofModel1}
 \eeq
 \beq{}
 \lambda'=\lambda\Phi'=\ell^{-1}\beta e^{-\frac{4}{3}\int_{-\infty}^\phi Xd\Phi}
 =-\ell^{-1}b\lambda^2 e^{\frac{4}{9}b\lambda}
 \\\Rightarrow
 \frac{du}{d\lambda}=(\lambda')^{-1}=-\frac{\ell e^{-\frac{4}{9}b\lambda}}{b\lambda^2}
 \;,\;\;\ell\equiv\frac{6}{\sqrt{3V_0}}
 \label{dudlambdaMod1}
 \eeq
so the dual gravity background can be written out explicitly
 \beq{}
 ds^2&=&e^{2A}(-dt^2+d\vec{x}\cdot
 d\vec{x})+\ell^2e^{2D}d\lambda^2
 \label{metricModel1}\\
 \Phi(\lambda)&=&\log\lambda
 \;,\;e^A=e^{\frac{1}{b\lambda}}\;,\;e^D=\frac{e^{-\frac{4}{9}b\lambda}}{b\lambda^2}
 \label{e2DofModel1}
 \eeq
where we use $\lambda$ as the radial holographic coordinate.
Using eq\eqref{VofPhiGeneral}, we know that the potential of
this model's dilaton field is,
 \beq{}
 V(\lambda)=V_0(1-\frac{b^2\lambda^2}{9})e^{\frac{8}{9}b\lambda}
 \eeq
From this potential, we easily see that the dual gravity background
is asymptotically AdS in the ultraviolet limit but not so in the
infrared region.

From equation \eqref{dudlambdaMod1}, we know that
 \beq{}
 \int_\infty^udu=-\ell\int_0^\lambda\frac{e^{-\frac{4}{9}b\lambda}}{b\lambda^2}d\lambda
 \eeq
By definition and iteration formulas of the incomplete Gamma
function, this leads to
 \beq{}
 u&=&\frac{4}{9}\ell\int_\lambda^\infty
 \frac{e^{-\frac{4}{9}b\lambda}}{(\frac{4}{9}b\lambda)^2}d(\frac{4}{9}b\lambda)
 =\frac{4}{9}\ell\cdot\Gamma\big[-1,\frac{4}{9}b\lambda\big]
 \nonumber\\
 &=&\frac{4}{9}\ell\cdot\Big[e^{-\frac{4}{9}b\lambda}\big(\frac{4}{9}b\lambda\big)^{-1}
 +\Gamma\big[0,\frac{4}{9}b\lambda\big]\Big]
 \label{holoCoor-u-mod1}
 \eeq
Note that $x\Gamma[0,x]_{x\rightarrow0}\rightarrow0$,
relative to the first term, the second term on the r.h.s of
\eqref{holoCoor-u-mod1} can be neglected. So in the ultraviolet
region, $\lambda\rightarrow0$, $u\rightarrow\frac{\ell}{b\lambda}$, $A=\frac{1}{b\lambda}\rightarrow\frac{u}{\ell}$. Using domain
wall coordinates \eqref{domainWall-bgrnd}, the gravity background
can be easily looked out being asymptotically AdS. While in the infra-red
region $\lambda\rightarrow\infty$, the scale factor of the domain
wall coordinate $e^A=e^{\frac{1}{b\lambda}}\rightarrow1$ this forms
a natural softwall for the dual gravity background.

For model 2, by the same procedure as previous we know
the dual gravity background and the corresponding dilaton
potential are
 \beq{}
 \beta&=&-b_0\lambda^2-b_1\lambda^3=\frac{d\lambda}{dA}
 \nonumber\\
 \Rightarrow ds^2&=&e^{2A}(-dt^2+d\vec{x}\cdot
 d\vec{x})+\ell^2e^{2D}d\lambda^2
 \label{metricModel2}\\
 \Phi(\lambda)&=&\log\lambda
 \;,\;e^A=e^{\frac{1}{b_0\lambda}}\Big[\frac{b_0}{b_1\lambda}+1\Big]^{-\frac{b_1}{b_0^2}}
 \nonumber\\
 &&\phantom{\log\lambda}e^D=\frac{e^{-\frac{4}{9}(b_0\lambda+\frac{1}{2}b_1\lambda^2)}}{b_0\lambda^2+b_1\lambda^3};
 \label{e2A-e2D-model2}
 \eeq
and
 \beq{}
 V(\lambda)&=&V_0[1-\frac{1}{9}(b_0\lambda+b_1\lambda^2)^2]
 e^{\frac{8}{9}(b_0\lambda+\frac{1}{2}b_1\lambda^2)}
 \eeq
respectively.
Obviously, in the ultraviolet region $\lambda\rightarrow0$,
this model has the same asymptotical background ---AdS space-time --- as model 1.
While in the infra-red region,
 \beq{}
 \frac{du}{d\lambda}&\rightarrow&-\ell\frac{e^{-\frac{2}{9}b_1\lambda^2}}{b_1\lambda^3}
 \nonumber\\
 \Rightarrow u&=&\frac{4}{9}\ell\cdot\Gamma\big[-1,\frac{2}{9}b_1\lambda^2\big]
 \nonumber\\
 &=&\frac{4}{9}\ell\cdot\Big[e^{-\frac{2}{9}b_1\lambda^2}\big(\frac{2}{9}b_1\lambda^2\big)^{-1}
 +\Gamma\big[0,\frac{2}{9}b_1\lambda^2\big]\Big]
 \eeq
So in this model, the rate of
$u\rightarrow0$ as $\lambda\rightarrow\infty$ is different from
that in model 1, but the qualitative conclusions that, (i) $u\rightarrow0$
as $\lambda\rightarrow\infty$, (ii) the dual gravity background
is not asymptotically AdS in the infra-red limit, (iii) the dual
gravity space-time's scale factor $e^A_{\lambda\rightarrow\infty}\rightarrow1$
are common between the two models.

For model 3,
 \beq{}
 \beta&=&-\frac{b_0\lambda^2}{1+b_1\lambda}=\frac{d\lambda}{dA}
 \nonumber\\
 \Rightarrow ds^2&=&e^{2A}(-dt^2+d\vec{x}\cdot
 d\vec{x})+\ell^2e^{2D}d\lambda^2
 \label{metricModel3}\\
 \Phi(\lambda)&=&\log\lambda\;,\;\;
 e^A=e^{\frac{1}{b_0\lambda}}(b_0\lambda)^{-\frac{b_1}{b_0}}
 \nonumber\\
 &&\phantom{\log\lambda}
 e^D=\frac{(1+b_1\lambda)^{1-\frac{4b_0}{9b_1}}}{b_0\lambda^2}
 \label{e2A-e2D-model3}
 \\
 V(\lambda)&=&V_0[1-\frac{b_0^2\lambda^2}{9(1+b_1\lambda)^2}](1+b_1\lambda)^{\frac{8b_0}{9b_1}}
 \eeq
In the ultraviolet region, $\lambda\rightarrow0$,
$\frac{du}{d\lambda}=-e^D\rightarrow-\frac{L}{b_0\lambda^2}$,
$u\rightarrow\frac{\ell}{b_0\lambda}$, $e^A\rightarrow{e}^{\frac{u}{L}}$.
So, like the previous two models, this model also has
asymptotically AdS geometry. But in the infra-red
region, this model's dual space-time scale factor
$e^A\rightarrow(b_0\lambda)^{-\frac{b_1}{b_0}}$ shrinks to zero
as $\lambda\rightarrow\infty$, the corresponding domain wall
coordinate $u$ approaches zero as $\lambda\rightarrow\infty$
by the following law
 \beq{}
 u&\sim&\ell\frac{9b_1^2}{4b_0^2}(b_1\lambda)^{-\frac{4b_0}{9b_1}}.
 \eeq

In summary, all three models have the same ultraviolet asymptotic
geometries --- AdS space-time plus running dilaton field. In the infrared
region, the first two model's dual space-time scale factor approaches to
1 as $\lambda\rightarrow\infty$,
but third model's scale factor shrinks to zero size as
$\lambda\rightarrow\infty$. The domain wall coordinate $u$ of model
1 approaches zero at the rate $u\sim 2e^{-\frac{4}{9}b\lambda}$,
that of model 2 at the rate $u\sim 4e^{-\frac{2}{9}b_1\lambda^2}$,
while that the model 3, $u\sim(b_1\lambda)^{-\frac{4b_0}{9b_1}}$.

 \section{Heavy quark potentials}
The heavy quark potential is an important quantity related with
confinement. It is measured with high precision in lattice
simulations, see reference \cite{QCDforces0001} and
\cite{ConellPotential}. The results are usually fit into
rational polynomials
\beq{}
 &&E(\rho)=-\frac{\kappa}{\rho}+\frac{\kappa'}{\rho^2}+\frac{\rho}{a^2}+C
 \label{conellPotential}
 \eeq
where $\rho$ is the distance between the quark and anti-quark
while $\kappa$, $\kappa'$ and $a^2$ are
parameters to be determined by fittings, $C$ is an irrelevant normalizing
constant. In gauge theories, this potential is related to
the expectation value of rectangular Wilson loops
of width $\rho$ and length $T$
\beq{}
\langle W\rangle=\exp[-E(\rho)T]
\eeq
here $T$ can be understood as the time
the quark and anti-quark are bounded together. By the gauge/string duality conjectures,
$\langle W\rangle$ equates to the partition function of an open string
which is living on the AdS background and whose world sheet boundary coincides
with the Wilson loop
 \beq{}
 \langle W\rangle=\exp[-E(\rho)T]\sim Z=\int\mathcal{D}X^\mu\exp(-S)
 \label{wLoopExpectConjecture}
 \eeq
where $S$ is the Nambu-Goto action of the string. Let $x^\mu(\sigma,\tau)$
denote the classical string profile and $\xi^\mu$ the quantum fluctuation around it.
Obviously, the path integral will be dominated by the classical
string configuration since it minimizes $S$, this means
\beq{}
Z&=&\exp(-S_c[x])\int\mathcal{D}\xi\exp{[-S(x+\xi)+S_c(x)]}
\nonumber\\
&\approx&\exp(-S_c[x])\int\mathcal{D}\xi\exp{[-\xi^{a\dagger}\mathcal{O}_a\xi^a]}
\label{partitionFuncToqquantum}
\eeq
So in the classical approximation, the heavy quark potentials read
\beq{}
E(\rho)T\approx S_c[x]
\eeq
While to first order quantum corrections
\beq{}
E(\rho)T\approx S_c[x]-\sum\log\det\mathcal{O}_a
\label{potentialsQQuantum}
\eeq
In the following, we will calculate the heavy
quark potentials quantitatively in the classical approximations and
study the quantum corrections qualitatively. We do not
get quantitative results on the latter. But even in
the classical approximation level, the models provided
in this paper predict heavy quark potentials more consistently
fitting with the lattice QCD results than almost all the models
considered in \cite{comparingPotentials-adsQCD}.

\subsection{Classical Approximations}

In the classical approximations, to have the given
Wilson loop as the world sheet boundary, the
string must have its two ends fixed on the boundary
and the center part dips into the bulks of the background AdS space-time
 \begin{subequations}
 \beq{}
 x^0=\tau\;,\;x^1=\sigma\;,\;\lambda=\lambda(\sigma)\;,\;x^{2,3}\;\mathrm{fixed}
 \\
 \sigma\in\Big[-\frac{\rho}{2}\;,\;\frac{\rho}{2}\Big]
 \\
 \lambda(-\frac{\rho}{2})=\lambda(\frac{\rho}{2})=0
 \;,\;\lambda(0)=\lambda_0
 \eeq
 \end{subequations}
For this configuration, the classical Nambu-Goto action can be written as
$S_c=E_c\cdot\int d\tau$, with
 \beq{}
 E_c=\frac{1}{\pi\alpha'}\int_0^{\frac{\rho}{2}}
 d\sigma\sqrt{e^{\frac{8\Phi}{3}}e^{2A}(e^{2A}+\ell^2e^{2D}\lambda_{,\sigma}^2)}
 \label{qqbarEnergy-unsubtract}
 \eeq
where $\lambda_{,\sigma}$ denotes $\frac{d\lambda}{d\sigma}$.
Since the classical configuration minimizes $S_c$ we know
 \beq{}
 \frac{e^{\frac{8\Phi}{3}}e^{4A}}
 {\sqrt{e^{\frac{8\Phi}{3}}e^{2A}(e^{2A}+\ell^2e^{2D}\lambda_{,\sigma}^2)}}
 =\mathrm{const.}=e^{\frac{4}{3}\Phi_0}e^{2A_0}
 \label{dlambdaDsigmaEOM}
 \\
 \Rightarrow\,\ell\lambda_{,\sigma}=\pm e^{A-D}\sqrt{e^{\frac{8}{3}\tilde{\Phi}}e^{4\tilde{A}}-1}
 \label{dlambdaDsigmaSOL}
 \eeq
in the above derivation, we have used the fact that at the center point of the string
$\sigma=0$, $\lambda_{,\sigma}=0$.
Using eqs\eqref{dlambdaDsigmaEOM} and \eqref{dlambdaDsigmaSOL},
both the distance $\rho$ between the two ends of the string and
its energy $E_c$ can be expressed as functions of $\lambda_0$.
 \beq{}
 \rho&=&2\ell e^{-A_0}\int_0^{\lambda_0}d\lambda
 \frac{e^{D-3\tilde{A}}\tilde{\lambda}^{-\frac{4}{3}}}
 {\sqrt{1-\tilde\lambda^{-\frac{8}{3}}e^{-4\tilde{A}}}}
 \label{qqbarDistance}
 \\
 E_c&=&\frac{\ell\lambda_0^{\frac{4}{3}}e^{A_0}}{\pi\alpha'}
 \int_0^{\lambda_0}d\lambda
 \frac{\tilde{\lambda}^{\frac{4}{3}}e^{D+\tilde{A}}}
 {\sqrt{1-\tilde\lambda^{-\frac{8}{3}}e^{-4\tilde{A}}}}
 \label{qqbarEnergy-unsubtract2}
 \eeq
where $\tilde{A}\equiv A-A_0$ while $\tilde{\lambda}=\frac{\lambda}{\lambda_0}$.
Since the string is infinitely long, the total energy in \eqref{qqbarEnergy-unsubtract2}
is divergent. In the Gauge theory side, this means that quark-antiquark pairs
are infinitely heavy. To obtain the finite interaction energy, we will subtract from
$E_c$ the infinite rest mass part, which corresponds to that of two straight
strings, see reference \cite{wilsonLoop1} and \cite{wilsonLoop2}.
Taking model 1 as an example, this means
 \beq{}
 E_{c,q\bar{q}}&=&\frac{\ell\lambda_0^{\frac{4}{3}}e^{A_0}}{\pi\alpha'}
 \Bigg[\int_0^{\lambda_0}d\lambda
 \frac{\tilde{\lambda}^{\frac{4}{3}}e^{D+\tilde{A}}
 \big[1-\sqrt{\cdots}\big]}
 {\sqrt{1-\tilde{\lambda}^{-\frac{8}{3}}e^{-4(\tilde{\lambda}^{-1}-1)/c}}}
 \nonumber\\
 &&\rule{3cm}{0pt}
 -\int_{\lambda_0}^{\infty}{d\lambda}{\tilde{\lambda}^{\frac{4}{3}}e^{D+\tilde{A}}}
 \Bigg]
 \label{qqbarPotentialMod1-1}
 \eeq
where $\tilde\lambda\equiv\frac{\lambda}{\lambda_0}$,
$c{\equiv}b\lambda_0$ and the `$\sqrt{\cdots}$' still means an expression
the same as the denominator. To assure the integrand of
the right hand side take real values in the range
$\tilde{\lambda}\in[0,1]$, the parameter $c$ must be less
than $\frac{3}{2}$. Through qualitative analysis of, we know that
$\rho$ is a monotonically increasing function of $c$.

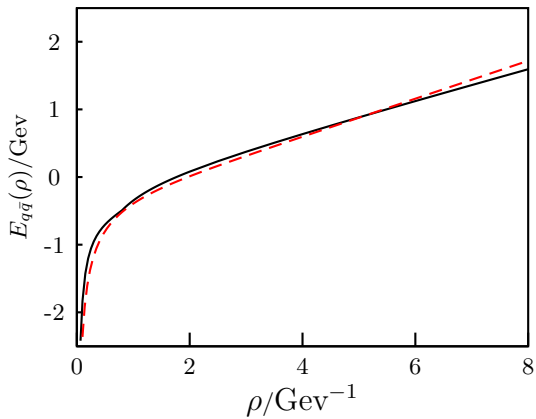
\begin{figure}[h]
 \psset{xunit=0.75cm,yunit=0.9cm}
 \begin{pspicture}(-1.6,-3.5)(8,2.5)
 \psaxes[axesstyle=frame,Ox=0,Oy=-2.5,ticks=none]{->}(0,-2.5)(8,2.5)
 \multido{\nx=0+2}{5}{\psline(\nx,-2.5)(\nx,-2.3)\rput(\nx,-2.8){\nx}}
 \multido{\ny=-2+1}{5}{\psline(0,\ny)(0.1,\ny)\rput(-0.4,\ny){\ny}}
 \rput{90}(-1,0){$E_{q\bar{q}}(\rho)/$Gev}
 \rput(4,-3.3){{\large$\rho$}{\small/}{\large Gev$^{-1}$}}
 \listplot[yStart=-2.5,xEnd=8]{\rhoEqqbAnaAprxMone}
 \listplot[xunit=4,linecolor=red,linestyle=dashed,xStart=0.02,xEnd=2]{\rhoEqqbListaMone}
 \end{pspicture}
 \caption{Heavy quark potentials in model 1, the continuous
 line is the result of analytical approximation
 \eqref{qqbarPotential-mod1}, while the dashed line comes from
 direct numerical integration of eq\eqref{qqbarDistance}
 and \eqref{qqbarPotentialMod1-1}.
 }
 \label{figurePotential-mod1}
 \end{figure}

Under the long separation limit which corresponds to $c\rightarrow\frac{3}{2}$,
the potentials can be calculated easily
 \beq{}
 E_{c,q\bar{q},\rho\rightarrow\infty}
 &\stackrel{c\rightarrow\frac{3}{2}}{\approx}&
 \frac{\lambda_0^\frac{4}{3}e^{2A_0}}{2\pi\alpha'}\cdot\rho
 +\frac{\ell\lambda_0^{\frac{4}{3}}e^{A_0}}{\pi\alpha'}
 \int_{\lambda_0}^{\infty}{d\lambda}{\tilde{\lambda}^{\frac{4}{3}}e^{D+\tilde{A}}}
 \nonumber\\
 &=&\frac{\lambda_0^\frac{4}{3}e^{2A_0}}{2\pi\alpha'}\cdot\rho
 +\frac{\ell \big(\frac{3}{2}\big)^{\frac{2}{3}}e^{\frac{1}{c}}}{\pi\alpha'b^\frac{4}{3}}
 \Gamma\big[\frac{1}{3},\frac{4c}{9}\big]_{c\rightarrow\frac{3}{2}}
 \label{qqbarPotentialMod1-L1}
 \eeq
While at the short separation limit $c\rightarrow0$.
 \beq{}
 \rho&\stackrel{c\rightarrow0}{\approx}&
 2\ell e^{-\frac{1}{c}}(0.596+0.0471c+\mathcal{O}[c^2])
 \label{qqbarDistanceMod1-S2}
 \\
 E_{c,q\bar{q}}
 &\stackrel{c\rightarrow0}{\approx}&
 \frac{2\ell c^{\frac{4}{3}}e^{\frac{1}{c}}}{\pi\alpha'b^\frac{4}{3}}(-0.602+0.042c+\mathcal{O}[c^2])
 \label{qqbarPotenMod1-S2}
 \eeq
Which means that
 \beq{}
 E_{c,q\bar{q},\rho\rightarrow0}&=&
 -\frac{2\ell^2}{\pi\alpha'b^{4/3}}
 \frac{\mathcal{\epsilon}(c)|_{c=c(\rho)}}{\rho}
 \nonumber\\
 \epsilon(c)&=&c^{4/3}
 \Big[0.359+0.003c+\cdots\Big]
 \label{qqbarPotentialMod1-S3}
 \eeq
the functional form of $c(\rho)$ should be obtained by
reversing the relation \eqref{qqbarDistanceMod1-S2}.
For example, to first order approximation in
$\big[\log\frac{\ell}{\rho}\big]^{-1}$, we can use iteration method to obtain
 \beq{}
 c\approx\Big(\log\big[\frac{1.19\ell}{\rho}
 +\frac{0.0942\ell}{\rho}\big(\log\frac{1.19\ell}{\rho}\big)^{-1}\big]\Big)^{-1}
 \label{cofEpsilonFirstOrder}
 \eeq
Since the series expression for $\rho(c)$
in eq\eqref{qqbarDistanceMod1-S2} is incomplete, its inverse
\eqref{cofEpsilonFirstOrder} is not well-defined
as $\rho>1.19\ell$. We assume that
$c(\rho)_{\rho>0.8/\textrm{Gev}}\equiv c(0.8/\textrm{Gev})$.
Combining eqs\eqref{qqbarPotentialMod1-L1} and
\eqref{qqbarPotentialMod1-S3} together,
the classically approximated heavy quark potential of model 1 can be written as follows
 \beq{}
 E_{c,q\bar{q}}(\rho)=-\frac{2\ell^2}{\pi\alpha'b^{4/3}}\frac{\epsilon(c)_{c=c(\rho)}}{\rho}
 +\frac{\big(\frac{3}{2}\big)^\frac{4}{3}e^\frac{4}{3}}{2\pi\alpha'b^\frac{4}{3}}\rho
 \nonumber\\
 -\frac{\big(\frac{3}{2}\big)^{\frac{2}{3}}e^{\frac{2}{3}}\ell}{\pi\alpha'b^{\frac{4}{3}}}
 \Gamma\big[\frac{1}{3},\frac{2}{3}\big]
 \label{qqbarPotential-mod1}
 \eeq
Figure \ref{figurePotential-mod1} compares this
result with the direct numerical integrations intuitively.
From the figure we see that the analytical formula captures
the main properties of the integration remarkably well.
For model 2 and model 3, the potentials can be calculated
similarly, and the results are also similar. But the precision
of the analytical approximation is lower than model 1. For
this reason, in the following fittings, we use numerics
to get the $\rho-E_{c,q\bar{q}}$ relations for all the three
models directly.

Eq\eqref{qqbarPotential-mod1} does not have the
the standard Cornell potential form, for its $\frac{1}{\rho}$
term(or the Luscher term) has a $\rho$-dependent coefficient,
which is proportional to $[\log\frac{1.19\ell}{\rho}]^{-\frac{4}{3}}$.
This is determined by asymptotical behavior
\footnote{If we assume that
\beq{}
\beta\rightarrow-b\lambda^p\,,\;p>1
\eeq
then, by the same calculations as above, we will find that
\beq{}
E_{q\bar{q},\rho\rightarrow}=-\frac{[\log\frac{1.19\ell}{\rho}]^{-\frac{4}{3(p-1)}}}{\rho}
+\mathrm{other\,terms}
\eeq} of the beta function
$\beta\propto -\lambda^2$ and is a common prediction of all the
renormalization group revised models constructed by
GKN's general procedure \cite{GKN0707-1324}, \cite{GKN0707-1349}.
Just as we will show in the following, this feature makes these models
predict $\rho-E_{q\bar{q}}$ relations more consistently
fitting with the lattice data than those which do not
include the renormalization effects.

\subsection{Comparing with lattice results}
\begin{figure}[h]
 \psset{xunit=0.6cm,yunit=0.9cm}
 \begin{pspicture}(-2,-3.5)(10,2.5)
 \psaxes[axesstyle=frame,Ox=0,Oy=-2.5,ticks=none]{->}(0,-2.5)(10,2.5)
 \multido{\nx=0+2}{6}{\psline(\nx,-2.5)(\nx,-2.3)\rput(\nx,-2.8){\nx}}
 \multido{\ny=-2+1}{5}{\psline(0,\ny)(0.2,\ny)\rput(-0.4,\ny){\ny}}
 \rput{90}(-1.2,0){$\big[E_{q\bar{q}}(\rho)-E_{q\bar{q}}(\rho_0)\big]/$Gev}
 \rput(5,-3.3){{\large$\rho$}{\small/}{\large Gev$^{-1}$}}
 \dPlotEbar{\dataLatticeGG}
 {\psset{xunit=5.0684}
 \listplot[linecolor=red,xStart=0.02,xEnd=1.973]{\rhoEqqbListaMone}
 \listplot[linecolor=green,xStart=0.02,xEnd=1.973]{\rhoEqqbListaMtwo}
 \listplot[linecolor=blue,xStart=0.04,xEnd=1.973]{\rhoEqqbListaMthree}
 }\end{pspicture}
 \caption{Potentials $E_{q\bar{q}}(\rho)$ predicted by our
 AdS/QCD models. The discrete data points comes from the
 lattice results of quenched QCD, reference
 \cite{heavyquarkPotential-lattice}. Since the three models
 almost fit the lattice data equally well, the three
 lines coincide.
 }\label{figCompareWithLattice}
 \end{figure}
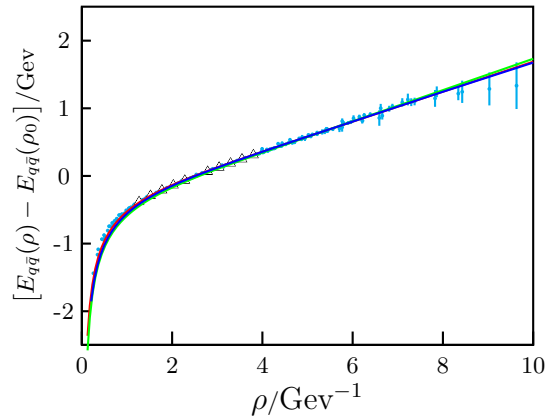

Figure \ref{figCompareWithLattice} compares the
the heavy quark potentials following from our AdS/QCD
models and those of lattice results. The relevant
parameters are listed in table \ref{tableParameters}
and are fixed by fits, which is required to minimize the
following $\chi^2$ variable
\beq{}
\chi^2=\sum_{i=1}^{N_{data}}\frac{(E_i^{AdS/QCD}-E_i^{latt.})^2}{\sigma_i^2}
\eeq
where the sum is over all the used data points and
$\sigma_i$ is the absolute error of the ith point.
\begin{table}
 \begin{tabular}{llll}
 \hline
 \rule{1.5cm}{0pt} & $\rho_{cut}$\rule{5mm}{0pt} & parameters\rule{1cm}{0pt} & Best fit\\
 \hline
 Model\,1 & 1.0 & $\alpha'b_0^\frac{4}{3}$ & \;4.99\\
 \;&\;& $\ell$ & \;4.71\\
 \;&\;& const & -0.013\\
 Model\,2 & 1.0 & $\alpha'b_0^\frac{4}{3}$ & \;4.38\\
 \;&\;& $\ell$ & \;4.37\\
 \;&\;& $b_1/b_0^2$&\;0.011\\
 \;&\;& const &\;0.018\\
 Model\,3 & 1.0 & $\alpha'b_0^\frac{4}{3}$ &\;4.55\\
 \;&\;& $\ell$ &\;4.25\\
 \;&\;& $b_1/b_0$&\;0.058\\
 \;&\;& const & -0.029\\
 \hline
 \end{tabular}
 \caption{Best fit values for the parameters involved in
 our AdS/QCD models. All dimensional parameters use Gev or
 Gev$^{-1}$ as units.}
 \label{tableParameters}
 \end{table}
 From the figure we see that almost all the three models
 fit with the lattice data equally well. This means that
 in the AdS/QCD framework, only by fitting with lattice
 data on heavy quark potentials, we cannot distinct the
 infrared behavior of the gauge theory
 $\beta$ function.

\begin{figure}[h]
 \psset{xunit=1.2cm,yunit=0.7714cm}
 \begin{pspicture}(-1,-1.1)(5,5.9)
 \psaxes[axesstyle=frame,Ox=0,Oy=0,ticksize=6pt]{->}(0,0)(5,5.9)
 \rput{90}(-0.7,3){\large$\chi^2$/d.o.f}
 \rput(2.5,-0.8){\large$\rho_{cut}$\small/Gev$^{-1}$}
 \listplot[xStart=0.85,linecolor=red,showpoints=true]{\chispdofMone}
 \listplot[xStart=0.85,linecolor=green,showpoints=true]{\chispdofMtwo}
 \listplot[xStart=0.85,linecolor=blue,showpoints=true]{\chispdofMthree}
 \psline[linestyle=dashed,dash=2pt 2pt](0,1)(5,1)
 \end{pspicture}
 \caption{Variation of the minimal $\chi^2/N_{d.o.f}$
 with respect to $\rho_{cut}$. The red, green and blue lines
 are the results for our model 1, model 2 and model 3 respectively.}
 \label{figChi2perdof}
 \end{figure}
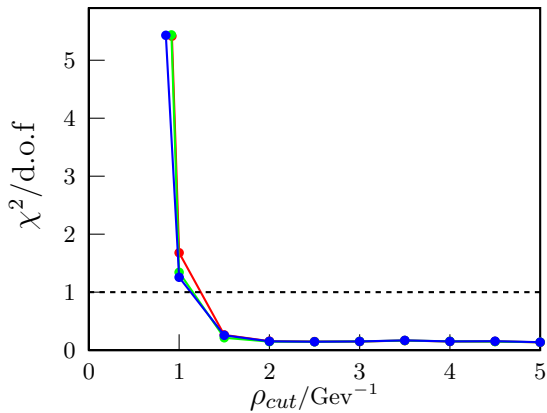

 Taking our FIG. \ref{figCompareWithLattice} and the Fig.1
 of reference \cite{comparingPotentials-adsQCD} as a
 comparison, we easily see that our models can be
 more consistently fitted with the lattice data than almost all
 the previous models which do not include the
 renormalization group improvements. To make this point more
 clearly, we display the $\chi^2/N_{d.o.f}\,-\,\rho_{cut}$
 in figure \ref{figChi2perdof}. Our $\rho_{cut}$ has the same
 meaning as that of reference \cite{comparingPotentials-adsQCD},
 which is the length scale below which the lattice data is
 not used to make fits. This is a manufacturing selection. Our
 models can give very low $\chi^2/N_{d.o.f}$ even when the
 $\rho_{cut}$ is set as low as $1$Gev$^{-1}$, while most of
 the AdS/QCD models which do not include the renormalization
 group effects can only do so when $\rho_{cut}$ is set as
 $4-5$Gev$^{-1}$. In a another word, to make a good fit,
 the usual models must cut more than one half of the lattice
 data points while our models almost need no cut at all.

 This is a really exciting result and it can be attributed
 to the fact that our models predict
 heavy quark potentials which at the short separation limit
 behaves like $E_{q\bar{q},\rho\rightarrow0}\rightarrow
 -[\ln\frac{1.19\ell}{\rho}]^{-\frac{4}{3}}/\rho$. Considering
 the fact that almost all the AdS/QCD models which do not
 include the renormalization group effects predict
 $E_{q\bar{q},\rho\rightarrow0}\rightarrow1/\rho$ and the
 quantum corrections do not change it remarkably\cite{wilsonLoop2}, a natural
 question arises, will the quantum corrections in our models
 change the short separation limit of our potentials remarkably?
 We change in the following to the quantum corrections
 of the heavy quark potentials.

\subsection{Quadratic quantum corrections}
Including quantum corrections, to first first order
approximation, the heavy quark potentials can be read out from
eqs\eqref{partitionFuncToqquantum} and \eqref{potentialsQQuantum},
\beq{}
E(\rho)T=S_c[x]-\sum_a\log\big(\det\mathcal{O}_a\big)
\label{totalEnergy-qqbar}
\eeq
To get quantitative results, we need to run 3 steps \cite{wilsonLoop2}.
(i) choose an appropriate gauge and define the fluctuations
properly, (ii) work out the corresponding action $S[x_c+\xi]-S[x_c]$
describing the fluctuations and read out from it the relevant
second order differential operators $\mathcal{O}_a$, (iii) calculate
eigenvalues of these operators and use some regularization
procedures to calculate the functional determinants of them
and finally extract the resulting corrections to the potential.

We will consider the fluctuations in the fixed $\lambda$
--- fix it on the classical profile --- gauge. In this gauge
the fluctuating quantities are $\xi$ and $\eta$, whose boundary
properties and relations to the complete embedding
functions $X^\mu$ are
\begin{subequations}
\beq{}
&&\hspace{-3mm}X^0=\tau\;,X^1=\sigma+\xi(\tau,\sigma)
\;,\;\;\lambda=\lambda(\sigma)\nonumber\\
&&\hspace{-3mm}\phantom{X^0=\tau\;,}X^{2,3}=\eta^{2,3}(\tau,\sigma)\;
\\
&&\hspace{-3mm}\xi(\tau,\sigma)_{\tau=0,T}=0\;,\;\;
\xi(\tau,\sigma)_{\sigma=0,\pm\rho/2}=0
\\
&&\hspace{-3mm}\eta(\tau,\sigma)_{\tau=0,T}=0\;,\;\;
\eta(\tau,\sigma)_{\sigma=0,\pm\rho/2}=0
\eeq
\end{subequations}
where $\lambda(\sigma)$ describes the classical string
profile which satisfies eq\eqref{dlambdaDsigmaSOL}.
Substituting these embeddings into eq\eqref{partitionFuncToqquantum}
and extract terms quadratic in quantities $\xi$ and $\eta$,
we have
\beq{}
S_{(2)}&=&\frac{1}{2}[G^s_{tt}(G^s_{xx}+G^s_{\lambda\lambda}\lambda'^2)]^{-\frac{1\,}{2}}
\Big[G^s_{\lambda\lambda}G^s_{xx}\lambda'^2\dot\xi^2+\nonumber\\
&&\hspace{-11mm}G^s_{tt}G^s_{xx}\xi'^2+G^s_{yy}(G^s_{xx}
+G^s_{\lambda\lambda}\lambda'^2)\dot\eta^2
+G^s_{tt}G^s_{yy}\eta'^2
\Big]\label{S2fxLambdaGauge}
\eeq
where the over-dots denote derivatives with respect to
$\tau$ and primes denote derivatives with respect to $\sigma$,
After integrating by parts and simplifying by
the classical e.o.m \eqref{dlambdaDsigmaEOM},
$S_{(2)}$ can be written in the form
$\xi^\dagger\mathcal{O}_\xi\xi+\eta^\dagger\mathcal{O}_\eta\eta$, with
\beq{}
\mathcal{O}_\xi&=&-(\lambda_0^\frac{4}{3}e^{2A_0})
[\frac{1}{\rho^2}\partial_{\tilde x}^2+(\tilde\lambda^\frac{8}{3}e^{4\tilde A}-1)\partial_t^2]
\label{Oxi-fxLambda-2}\\
\mathcal{O}_\eta&=&-(\lambda_0^\frac{4}{3}e^{2A_0})
[\frac{1}{\rho^2}\partial_{\tilde x}^2+\tilde\lambda^\frac{8}{3}e^{4\tilde A}\partial_t^2]
\label{Oeta-fxLambda-2}
\eeq
where $\tilde x$ is the normalized $x=\tilde x\cdot\rho$,
this means that as $x$ varies in the range $0<x<\rho$, $\tilde x$
varies in the range $0<\tilde x<1$. Note that $\eta$ is a two component vector and
$\dot\eta^2$ actually means $\dot\eta^{(2)}\dot\eta^{(2)}+\dot\eta^{(3)}\dot\eta^{(3)}$,
and similarly $\eta'^2=\eta^{(2)}{'}\eta^{(2)}{'}+\eta^{(3)}{'}\eta^{(3)}{'}$.

\begin{figure}[h]
\centering
 \psset{xunit=4.7cm,yunit=0.4cm}
 \begin{pspicture}(0.1,-1)(1,8)
 \psaxes[Dx=0.2,Dy=2,Ox=0.2,Oy=1,ticksize=4pt,axesstyle=frame](0.2,1)(0.2,1)(1,8)
 \dataplot[plotstyle=curve,linecolor=red]{\FtvCurModela}
 \dataplot[plotstyle=curve,linecolor=green,linestyle=dashed,dash=3pt 3pt]{\FtvCurModelb}
 \dataplot[plotstyle=curve,linecolor=blue,linestyle=dashed,dash=6pt 6pt]{\FtvCurModelc}
 \rput[cc]{0}(0.6,-0.5){$v$}
 \rput[cc]{0}(0.8,7.5){$F_t(v)$}
 \psline[linecolor=red](0.25,6)(0.4,6)
 \rput[cc]{0}(0.51,6){$\rho=0.01\ell$}
 \psline[linecolor=green,linecolor=green,linestyle=dashed,dash=3pt 3pt](0.25,5.3)(0.4,5.3)
 \rput[cc]{0}(0.51,5.3){$\rho=0.10\ell$}
 \psline[linecolor=blue,linecolor=blue,linestyle=dashed,dash=6pt 6pt](0.25,4.6)(0.4,4.6)
 \rput[cc]{0}(0.51,4.6){$\rho=0.21\ell$}
 \end{pspicture}
 \begin{pspicture}(0.1,-1)(1,8)
 \psaxes[Dx=0.2,Dy=2,Ox=0.2,Oy=1,ticksize=4pt,axesstyle=frame](0.2,1)(0.2,1)(1,8)
 \dataplot[plotstyle=curve,linecolor=red]{\FtvSimplea}
 \dataplot[plotstyle=curve,linecolor=green,linestyle=dashed,dash=3pt 3pt]{\FtvSimpleb}
 \dataplot[plotstyle=curve,linecolor=blue,linestyle=dashed,dash=6pt 6pt]{\FtvSimplec}
 \rput[cc]{0}(0.6,-0.5){$v$}
 \rput[cc]{0}(0.8,7.5){$F_t(v)$}
 \psline[linecolor=red](0.25,6)(0.4,6)
 \rput[cc]{0}(0.51,6){$\rho=0.01\ell$}
 \psline[linecolor=green,linecolor=green,linestyle=dashed,dash=3pt 3pt](0.25,5.3)(0.4,5.3)
 \rput[cc]{0}(0.51,5.3){$\rho=0.10\ell$}
 \psline[linecolor=blue,linecolor=blue,linestyle=dashed,dash=6pt 6pt](0.25,4.6)(0.4,4.6)
 \rput[cc]{0}(0.51,4.6){$\rho=0.21\ell$}
\end{pspicture}
 \caption{Left, the function $F_t(v)$ appearing in the $\mathcal{O}_\eta$
 operator of eqs\eqref{Oeta-fxLambda-2}(write it into the form of
 \eqref{BAFoperator}), the function depends on
 the interquark distance $\rho$. Right,
 $F_t(v)$ for the simple $AdS_5/CFT_4$ correspondence,
 the function is independent of the interquark distance $\rho$}
 \label{figureFtv}
 \end{figure}
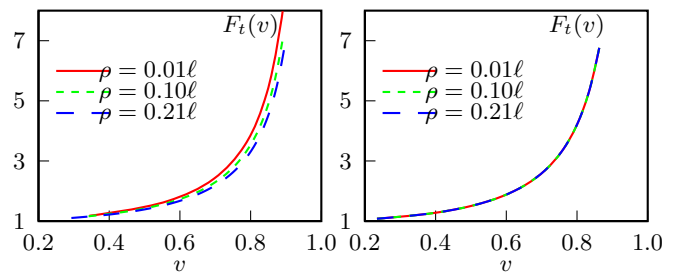

Reference \cite{wilsonLoop2}
proved that, for any operator of the form
\beq{}
\mathcal{O}[A,B]=B^2\mathcal{O}_v+A^2F_t(v)\partial_t^2
\label{BAFoperator}
\eeq
if the parameter $A$, $B$ encode all the dependences on
the inter-quark distance $\rho$, while $v$, $F_t$ and
$\mathcal{O}_v$ are variables
or operators independent of $\rho$, then the functional
determinant of $\mathcal{O}[A,B]$ must
be of the form
\beq{}
\log(\det\mathcal{O}[A,B])=\frac{BT}{A}\log(\det\mathcal{O}[1,1])+o(T)
\label{operatorDeterminant}
\eeq
i.e. $\rho$ dependence of the $\log\det$ of such operators
is completely contained in the coefficient $\frac{B}{A}$.
For almost all the models considered by \cite{wilsonLoop2},
say the typical $AdS_5/CFT_4$ case, this leads to that the
quadratic quantum correction to the heavy quark potentials are proportional to $\rho^{-1}$.
However, in the current case, if we write the operators
of eqs\eqref{Oxi-fxLambda-2} and \eqref{Oeta-fxLambda-2} into the form of eq\eqref{BAFoperator},
then the quantity $F_t(v)\equiv\tilde{\lambda}^\frac{8}{3}e^{4\tilde A}$
appearing in them will depend on the interquark distance implicitly,
see figure \ref{figureFtv} for details.

This is both a good and a bad news for us. The good news is,
it means that we can almost definitely conclude that the quantum corrections
to the heavy quark potentials is not of the $1/\rho$ type. The bad
news is, it means we cannot derive out explicit functional forms
for the $\rho$-dependence of the quadratic quantum corrections through
this simple method. To get quantitative results, we need to find out the eigenvalue
spectrum of $\mathcal{O}_\xi$ and $\mathcal{O}_\eta$ and their functional
determinant by numerics. Since they cannot be find out
analytically, the regularization of the resulting functional determinant
becomes a big problem. At this point, we have no good method
to overcome this obstacle numerically.

\begin{figure}
\centering
\psset{unit=6mm}
 \begin{pspicture}(-2,-5)(2,0)
 \psline(-2,0)(2,0)\rput(0,0.2){$\lambda=0$}
 \psplot[algebraic=true]{-0.7}{0.7}{4.898*(x*x-0.49)}
 \psline[linestyle=dashed](-2,-4)(2,-4)\rput(0,-3.7){$\lambda=\infty$}
 \put(0,-1){\psCoil[coilaspect=0,coilheight=0.8,coilwidth=0.3]{-790}{790}}
 \rput(0,-4.8){(a)}
 \end{pspicture}
 \rule{2mm}{0pt}\begin{pspicture}(-2,-5)(2,0)
 \psline(-2,0)(2,0)\rput(0,0.2){$\lambda=0$}
 \psplot[algebraic=true]{-0.7}{0.7}{4.898*(x*x-0.49)}
 \psline[linestyle=dashed](-2,-4)(2,-4)\rput(0,-3.7){$\lambda=\infty$}
 \rput(0,-4.8){(b)}
 \end{pspicture}
 \rule{2mm}{0pt}\begin{pspicture}(-2,-5)(2,0)
 \psline(-2,0)(2,0)\rput(0,0.2){$\lambda=0$}
 \psline[linestyle=dashed](-2,-4)(2,-4)\rput(0,-3.7){$\lambda=\infty$}
 \psline(-1.2,0)(-1.2,-4)\psline(1.2,0)(1.2,-4)
 \put(0,-2.4){\psCoil[coilaspect=0,coilheight=1.2,coilwidth=0.3]{-1200}{1200}}
 \rput(0,-4.8){(c)}
 \end{pspicture}
 \caption{String configurations contributing to the
 Wilson loop correlator. The upper lines denote the
 boundary of the asymptotically AdS space, while the lower dashed lines
 denote its center point. According to reference \cite{DebyeScreening},
 figure b) is the dominant contributor to the Wilson-Polyakov loop correlators at
 the short separation limit; figure c) is the dominator
 at the long separation limit; figure a) displays  quantum corrections
 due to the graviton-dilaton exchange process.}
 \label{figQuantumCorrect}
\end{figure}
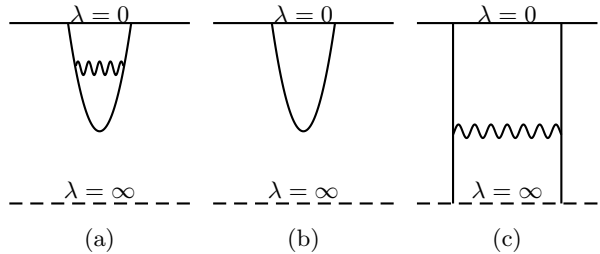

Reference \cite{DebyeScreening} studied the Wilson-Polyakov
loop correlators in the finite temperature $\mathcal{N}=4$ SYM
plasmas, from which the relevant heavy quark potentials can be
extracted conveniently. They pointed out that although at the
short separation limit, the heavy quark potentials is dominated
by smooth the string configuration which minimizes the classical
Nambu-Goto action, part(c) of FIG\ref{figQuantumCorrect},
at the long separation limit, the potentials
is dominated by string profiles which involves the exchange of
the lightest graviton-dilaton modes, see the part(c) of FIG.\ref{figQuantumCorrect}.
According to this results, our above considerations of the
quantum fluctuations(around the classical string profile) correction
is not the only contribution to the interquark-potentials. We should
also consider diagrams of the form like the part(a) of
FIG.\ref{figQuantumCorrect}. But this kind of string profile
contributing to the heavy quark potentials always carries a
factor of $\lambda^2$ relative to the classical profile does,
and to make the area of the string world sheet as small as possible,
the string cannot not fall very deep into the
bulk of the space-time, i.e. they are mainly around the $\lambda\rightarrow0$
region. So there is no need to worry that this kind of
correction will change the short separation limit of
our potentials.

 \section{Conclusions}
From the previous section, we know that the
heavy quark potentials following from all our three models
asymptote to the same form $E_{q\bar{q}}\propto-1/(\rho\big[\log\frac{1.19\ell}{\rho}\big]^{4/3})$
at the short separation limit. This is determined by
the UV property of the gauge theory beta function $\beta\propto-\lambda^2$
and is a common feature for all the renormalization group improved AdS/QCD models
constructed by GKN's procedure.
Although this is different from the usual Cornell potential,
through numerical fittings and comparison with reference
\cite{comparingPotentials-adsQCD}, we know that our
models can more consistently fit the lattice
data, relative to almost all the previous AdS/QCD models
which do not include the renormalization group effects,
for example, those of \cite{KKSS}, \cite{AndreevZakharov} and
\cite{bkreact-geometry-adsQCD}.

Numerical results also tells us that, though we can
construct AdS/QCD models with very different infrared $\beta$
function, for example $\beta\rightarrow-\lambda^2,-\lambda,-\lambda^3$,
only by comparison with the lattice QCD heavy quark potentials,
we cannot distinguish which one is the better.
Since at long separation limit we neglected the graviton-dilaton
exchange contributions to the heavy quark potentials,
we should be more careful on this conclusion. Just as reference
\cite{GKN0707-1349} showed and our numerics indicated, neglecting
the graviton-dilaton exchange contributions does not affect
the qualitative conclusion that the potentials following from our AdS/QCD
models are linearly confined. (Reference \cite{DebyeScreening} pointed
out that there is many uncertainties on the quantitative calculation
of the graviton-dilaton exchange contributions
at the long separation limit) Nevertheless, further
studies on this point and comparison with both the lattice data and
the heavy quark effective field theories
\cite{Quarkonium} will be interesting topics.

The above content consists the main conclusions of this paper.
On the model building part, we have no unexpected conclusions. A little
summary can be given as follows. Three AdS/QCD models are built by the
systematic procedure of GKN. At the perturbative limit, the first
model has gauge theory $\beta$ function coincide with the usual
QCD to 1 loop order; model 2's $\beta$ function coincides with the desiring
to 2 loops; while model 3, only reproduces the desiring to 1 loop order.
Nevertheless, due to non-perturbative effects, the $\beta$ function
of model 3 may be the one most closely related to the real QCD theory.
At the strong coupling limit, model 1's $\beta$ function approaches infinite as
$-\lambda^2$, model 2 as $-\lambda^3$, while model 3, $-\lambda$.
By GKN's general confinement criteria, model 1 and model 2 are
super-confining, while model 3 is critically confining.
In the ultraviolet limit, the three models' dual
gravity description have the same asymptotical background
--- AdS space-time plus running dilaton field. In the infrared region, model 1 and 2's
dual space-time scale factor approaches to 1 as $\lambda\rightarrow\infty$,
but model 3's scale factor shrinks to zero size as
$\lambda\rightarrow\infty$.

For future directions, we think looking for finite temperature
solutions to these models and the relevant phenomenological
applications of them may be interesting
subjects, see e.g. references \cite{finitTempAdSQCD-Herzog}
 \cite{finiteTempAdSQCD-AndreevZakharov}
 \cite{dragForceJetQuenching}
 \cite{finiteTempAdSQCD-GKMN}
 \cite{finiteTempAdSQCD-GNPR}
 \cite{chiralCondensate-holoQCD}
 \cite{thermoPhaseTrans-holoModels}. We hope to come back on this point in the near future.

\section*{Acknowledgements}
We thank very much to professor G. S. Bali and C. D. White for
reminding us their work, especially the former for
explaining the lattice data related questions,
including the normalization and obtaining conditions.

\appendix
\section*{Appendix}
\addcontentsline{toc}{section}{Appendix}
\numberwithin{equation}{section}
\renewcommand{\theequation}{\thesection.\arabic{equation}}

\section{Incomplete Gamma Function}\label{sectionIncompleteGammaFunc}

This appendix collects some basic formulas
of the incomplete Gamma function which is needed
in the analytical derivation of the short separation
limit of the heavy quark potentials. The content
is based on the help document of software $Mathematica$ and
some simple variable replacements and derivations.

The incomplete Gamma function $\Gamma[a,z]$ is defined as
 \beq{}
 \Gamma[a,z]=\int_z^\infty t^{a-1}e^{-t} dt
 \eeq
When $a>0$, we have the following expansion formulae,
 \beq{}
 \Gamma[a,z]&\stackrel{z\sim\infty}{=}&
 e^{-z}{z^{a-1}\Big( 1 + \frac{a-1}{z} + \frac{a^2-3a+2}{z^2} +
      {\mathcal{O}(\frac{1}{z})}^3 \Big) }
      \nonumber\\
 \label{Gammaaz-zgotoinfty}
 \\
 \Gamma[a,z]&\stackrel{z\sim0}{=}&\Gamma[a]+z^{a}
 \Big(-\frac{1}{a}+\frac{z}{1+a}-\frac{z^2}{2(2+a)}+\mathcal{O}(z^3)\Big)
 \nonumber\\
 \label{Gammaaz-zgotozero}
 \eeq
When $a<0$, by definition, we can easily prove
the following iteration formula,
 \beq{}
 \Gamma[a,z]&\stackrel{a<0}{=}&
 -\frac{e^{-z}z^a}{a}+\frac{1}{a}\Gamma[a+1,z]
 \\
 &=&-\frac{e^{-z}z^a}{a}-\frac{e^{-z}z^{a+1}}{a(a+1)}+\frac{1}{a(a+1)}\Gamma[a+2,z]
 \nonumber
 \eeq

For integrations like $\int_0^1
\lambda^{-p}e^{-\frac{1}{c\lambda}}d\lambda$, we can translate it
into incomplete Gamma functions through variable replacements
 \beq{}
 \int_0^1\lambda^{-p}e^{-\frac{1}{c\lambda}}d\lambda
 &=&\int_1^\infty
 \hat{\lambda}^{p-2}e^{-\frac{\hat{\lambda}}{c}}d\hat{\lambda}
 =c^{p-1}\Gamma\Big[p-1,\frac{1}{c}\Big]
 \nonumber\\
 \\
 \int_\epsilon^1
 \lambda^{-p}e^{-\frac{1}{c\lambda}}d\lambda
 &=&\int_1^\frac{1}{\epsilon}
 \hat{\lambda}^{p-2}e^{-\frac{\hat{\lambda}}{c}}d\hat{\lambda}
 \\
 &=&c^{p-1}\Gamma\Big[p-1,\frac{1}{c}\Big]
 -c^{p-1}\Gamma\Big[p-1,\frac{1}{c\epsilon}\Big]
 \nonumber
 \eeq

\end{document}